\documentclass{article}
\input{tcilatex}
\begin{document}

\begin{center}
\bigskip \textbf{TWO LOOP LOW TEMPERATURE CORRECTIONS TO ELECTRON SELF
ENERGY }

\bigskip Mahnaz Q. Haseeb$^{1^{\ast }}$ and\ Samina S. Masood$^{2}$

$^{1}$\textit{Department of Physics, COMSATS Institute of Information
Technology, Islamabad, Pakistan,}

$^{2}$\textit{Department of Physics, University of Houston Clear Lake,
Houston TX 77058}

$_{^{\ast }mahnazhaseeb@comsats.edu.pk}$

\bigskip

\textbf{Abstract}
\end{center}

We recalculate the two loop corrections in the background heat bath using
real time formalism. The procedure of the integrations of loop momenta with
dependence on finite temperature before the momenta without it, has been
followed. We determine the mass and wavefunction renormalization constants
in the low temperature limit of QED, for the first time with this preferred
order of integrations. The correction to electron mass and spinors in this
limit is important in the early universe at the time of primordial
nucleosynthesis as well as in astrophysics.

\bigskip



\section{Motivations\qquad \qquad}

The quantum field theory at finite temperature, Thermal Field Theory (TFT)
has been used for more than 50 years. The techniques for calculations in
many body systems were initially developed in condensed matter [1 ] which
are now also used to describe a large ensemble of multi-interacting
particles in a thermal background. Since the late seventies, TFT was used to
study the phase transitions in cosmology and quantum field theories. It was
adopted in particle physics to study the many particle systems. These
methods are used not only in particle physics but also study the various
aspects of nuclear matter and plasma physics. The TFT methods are
extensively used to describe the phase transitions due to symmetry breaking
after the Hot Big Bang and in tracing the history of the early universe.

The main applications of interest in this context are in:

\textit{Cosmology:} The early universe provides a very good example in the
studies of hot plasmas. When the Big Bang occurred, the Universe was
impossibly hot and dense. It rapidly expanded and cooled. At $t=200$ sec$,$ $%
T=10^{8}K,$ it was cool enough for neutrons and protons to combine to form
Deuterium, then Helium and traces of Lithium (primordial nucleosynthesis).
For the next few $10^{5}$ years it was too hot for electrons to form atoms.
The universe was filled with hot plasma of electrons and nuclei, bathed in
photons constantly interacting with both, like the interior of a star.

\textit{Astrophysics:} There is a series of different types of fusion
reactions in stars leading to luminous supergiants. When helium fusion
ceases in the core, gravitational compression increases the core's
temperature above $6\times 10^{8}K$ at which carbon can fuse into neon and
magnesium. As the core reaches $1.5\times 10^{8}K,$ oxygen begins fusing
into silicon, phosphorous, sulfur, and others. At $2.7\times 10^{8}K,$
silicon begins fusing into iron. This process essentially stops with the
creation of iron and a catastrophic implosion of the entire star initiates.
When the high mass stars exhaust their He fuel they have enough
gravitational energy to heat up to $6\times 10^{8}K.$ Cores of neutron
stars, red giants and white dwarfs are composed of extremely dense plasmas ($%
\rho =10^{6}$ -$10^{15\text{ }}g/cm^{3},)$. The neutrinos and axions
emission rates in these stars require TFT [2]. A tremendous amount of energy
is released in a supernova. The only supernova in modern time, visible to
the naked eye, was detected on Feb. 23, 1987 and is known as $SN1987A.$It
emitted more than $10^{10}$ times as much visible light as the Sun for over
one month and temperatures as high as $2\times 10^{11}K$ $\ $were reached.

\textit{Heavy Ion Collisions:} The quark gluon plasma is the form of matter
at transition temperatures $T_{c}=100-200$ $MeV$. The hot and dense
environment in quark gluon plasma and the studies of its prospective
reproduction in nucleus-nucleus collisions require the TFT methods for
detailed explanation. With the increased feasibility of creation of quark
gluon plasma in heavy ion collisions, the methods developed in this theory
got their specific relevance in QCD at finite temperature as well.\qquad

We are specifically interested in the first two applications here.

\section{Finite temperature effects\qquad \qquad \qquad \qquad \qquad}

The main idea of TFT is to use the approach of the usual quantum field
theory. Matsubara [3] was the first who developed thermal field theory by
incorporating a purely imaginary time variable in the evolution operator. In
Euclidean space the covariance breaks and time is included as an imaginary
parameter. The imaginary time domain is finite and periodic because of which
the energy integrations are converted into summations over the discrete
Matsubara frequencies. The presence of discrete energies along with the
particle distribution functions destroys the covariance of the theory.

The important contributions by Schwinger [4], Mills [5], and Keldysh [6] led
to the development of a formalism based upon the choice of a contour in the
complex plane. This is called the real time formalism. In the real-time
formalism, an analytical continuation of the energies along with Wick's
rotation restores covariance in Minkowski space at the expense of Lorentz
invariance. The breaking of Lorentz invariance leads to the non-commutative
nature of the gauge theories [7]. The covariance is incorporated through the
4-component velocity of the background heat bath $u^{\mu }=(1,0,0,0).$In a
heat bath the particles are in constant interaction with the thermal
surroundings. Implementing these interactions is straightforward as is done
in vacuum field theory. The temperature is included through the statistical
distribution functions of the particles.

Umezawa and coworkers [8] independently worked on a different approach
called Thermo- Field Dynamics that also gives the same results. In this
formalism, the propagators are taken in the form of $2\times 2$ matrices.
Field theory at finite temperature is renormalizable, if the vacuum theory
is so since the presence of the Boltzmann factor in the thermal corrections
cuts off any ultraviolet divergence. Choosing the suitable counter terms as
in vacuum can eliminate them. The infrared divergences are inherent in
almost all perturbation theories, whether at zero or finite temperature. KLN
theorem [9] demonstrates that singularities appearing at intermediate stages
of the calculation cancel in the final state physical result.

Quantum Electro Dynamics (QED) is the simplest and most successful gauge
theory. The behavior of QED at finite temperatures serves as a model for the
determination of background effects in other physical theories - the
electroweak theories as well as Quantum Chromo Dynamics. In the real time
formalism, the tree level fermion propagator in Feynman gauge in momentum
space is [10]

\begin{equation}
S_{\beta }(p)=(\NEG{p}-m)[\frac{i}{p^{2}-m^{2}+i\varepsilon }-2\pi \delta
(p^{2}-m^{2})n_{F}(E_{p})]\ ,
\end{equation}%
$\qquad $\ where

\begin{equation}
n_{F}(E_{p})=\frac{1}{e^{\beta (p.u)}+1}\ ,
\end{equation}%
\bigskip is the Fermi-Dirac distribution function with $\beta $=$\frac{1}{T}$%
. The boson propagator is taken as\bigskip

\begin{equation}
D_{\beta }^{_{\mu \nu }}(p)=[\frac{i}{k^{2}+i\varepsilon }-2\pi \delta
(k^{2})n_{B}(k)],
\end{equation}%
with

\begin{equation}
n_{B}(E_{k})=\frac{1}{e^{\beta (k.u)}-1}.
\end{equation}%
\bigskip

\section{One loop corrections}

At the one loop level, Feynman diagrams are calculated in the usual way by
substituting these propagators in place of the usual ones in vacuum. The
Lorentz invariance breaking and conserving terms remain separate at the one
loop level since the propagators comprise temperature dependent (hot) terms
added to temperature independent (cold) terms. This effect has been studied
in detail and established at the one-loop level [11]. The renormalization of
QED\ in this scheme [12] has already been checked in detail at the one loop
level for all temperatures and chemical potentials.

The thermal background effects are incorporated through the radiative
corrections. In finite temperature electrodynamics electric fields are
screened due to the interaction of the photon with the thermal background of
charged particles. The physical processes take place in a heat bath
comprising hot particles and antiparticles instead of vacuum. The exact
state of all these background particles is unknown since they continually
fluctuate between different configurations. The net statistical effects of
the background fermions and bosons enter the theory through the fermion and
boson distributions respectively.

The electric permittivity and the magnetic susceptibility of the medium are
modified by incorporating the thermal background effects. At low
temperatures, i.e., $T<<m_{e}$ ($m_{e}$ is the electron mass), the hot
fermions contribution in background is suppressed and only the hot photons
contribute from the background heat bath. The vacuum polarization tensor in
order $\alpha $ does not acquire any hot corrections from the photons in the
heat bath. This is because of the absence of self-interaction of photons in
QED.

The thermal mass is generated radiatively. The mass shift that enters
physical quantities acts as a kinematical cut-off, in the production rate of
light weakly coupled particles from the heat bath. The effective mass
corresponds to the fact that in the heat bath, the propagation of particles
is influenced by their continuous interactions with the medium.\bigskip

\section{Higher order corrections}

The higher order loop corrections are required to get predictions on
perturbative behavior at finite temperature. At the higher-loop level, the
loop integrals involve a combination of cold and hot terms which appear due
to the overlapping propagator terms in the matrix element. In such
situations, specific techniques are needed, even at the two loop level, to
solve them. Higher loops get analytically even more complicated. In the hot
terms there appear overlapping divergent terms. The removal of such
divergences is already shown at the two-loop level [13] for electron self
energy.

We restrict ourselves to the low-temperatures to prove the renormalizability
of QED at the two-loop level through the order by order cancellation of
singularities. The results have been shown [14] to depend on the order of
doing the hot and cold integrations. The justification of this specific
order is the fact that the temperature dependent part corresponds to the
contribution of real background particles on mass-shell and incorporates
thermal equilibrium. The breaking of Lorentz invariance changes these
conditions for the cold integrals. We have checked that the renormalization
can only be proven with the preferred order of integrations, i.e., if
covariant hot integrals are evaluated before the cold ones. At the higher
loop level the vacuum polarization contribution is non zero, even at low
temperature. The calculations are simplified if the temperature dependent
integrations are performed before the temperature independent ones. The
temperature independent loops can then be integrated using the standard
techniques of Feynman parametrization and dimensional regularization as in
vacuum [15].

The second order in $\alpha $ corrections to the electron self energy at low
temperature has been calculated to be equal to \bigskip 
\begin{equation}
\Sigma _{\beta }(p)=\Sigma _{\beta }(p,T=0)-\frac{\alpha ^{2}}{4\pi ^{3}}%
\left[ \frac{1}{\varepsilon }[(\NEG{p}+6m)I_{A}-2I]-3(\NEG%
{p}+4m)I_{A}-4\gamma _{\mu }I^{\mu }-\frac{8\pi T^{2}}{3m^{2}}(\NEG{p}-m)%
\right] 
\end{equation}%
where the singularity \bigskip 
\[
\frac{1}{\varepsilon }=\frac{1}{\eta }-\gamma -\ln {\LARGE (}\frac{4\pi \mu
^{2}}{m^{2}}{\LARGE )},
\]%
\ \ 

is the usual ultraviolet divergence in vacuum field theory in MS bar scheme
of renormalization with $\gamma $ as the Euler- Mascheroni constant and
\bigskip 
\begin{equation}
I_{A}=8\pi \int \frac{dk}{k}n_{B}(k),
\end{equation}
$\qquad \qquad \qquad \qquad \qquad \qquad \qquad \qquad \qquad \qquad $%
\begin{equation}
\frac{I^{0}}{E}=\frac{2\pi ^{3}T^{2}}{3E^{2}v}\ln \frac{1+v}{1-v},
\end{equation}

\begin{equation}
\frac{\mathbf{I.p}}{|\mathbf{p}|^{2}}=\frac{2\pi ^{3}T^{2}}{3E^{2}v^{3}}[\ln 
\frac{1+v}{1-v}-2v],
\end{equation}%
with $v=\frac{|\mathbf{p}|}{p_{_{0}}}$ . The divergences of the form $I_{A}$
and $\frac{1}{\varepsilon }$ in Eqs. (6) and (7) cancel on addition of the
appropriate counter terms. The finite contribution then leads to the
physical mass of electron:

\begin{center}
\begin{equation}
m_{phys}^{2}=m^{2}\left( 1+\frac{2\alpha \pi T^{2}}{3m^{2}}+\frac{4\alpha
^{2}T^{2}}{3m^{2}}\right) ,
\end{equation}%
\ \ 

\bigskip\ \ \ \ 
\end{center}

\section{Results}

The electron mass and wave function renormalization can be obtained from the
two loop self-energy for electrons calculated in the previous section. From
Eq. (10) the change in the electron mass due to finite temperature up to the
order $\alpha ^{2}$ relative to the cold electron mass is

\begin{equation}
\frac{\delta m}{m}=\frac{T^{2}}{m^{2}}\ \left( \alpha \pi +4\alpha
^{2}\right) .\ \ \ \ \ 
\end{equation}

\FRAME{ftbpF}{7.6761in}{4.3232in}{0pt}{}{}{Figure}{\special{language
"Scientific Word";type "GRAPHIC";maintain-aspect-ratio TRUE;display
"USEDEF";valid_file "T";width 7.6761in;height 4.3232in;depth
0pt;original-width 10.7747in;original-height 6.0511in;cropleft "0";croptop
"1";cropright "1";cropbottom "0";tempfilename
'LC9ZM700.wmf';tempfile-properties "XPR";}}

\begin{center}
Fig. 1\ \ \ 
\end{center}

The renormalizability of the self mass of electron is rechecked through the
order by order cancellation of singularities at both loop levels. It can
also be noted that the second order term is much smaller than the first
order term.

Using the standard procedure, the wave function renormalization constant
comes out to be\qquad

\begin{equation}
Z_{2}^{-1}\bigskip =1+\frac{\alpha }{4\pi }\left( 4-\frac{3}{\varepsilon }%
\right) -\frac{\alpha }{4\pi ^{2}}\left( I_{A}-\frac{I^{0}}{E}\right) -\frac{%
\alpha ^{2}}{4\pi ^{2}}\left( 3+\frac{1}{\varepsilon }\right) I_{A}+\frac{%
2\alpha ^{2}T^{2}}{3\pi ^{2}m^{2}}.
\end{equation}

In Fig. 1, a plot of $Z_{2}^{-1}$ vs $T/m$ is given for $O(\alpha ^{2})$\ to
demonstrate the affect of temperature in two-loops. The results here and in
Ref. [14] are an explicit proof of the renormalizability of QED up to the
two-loop level. They also estimate the temperature dependent

modification in the electromagnetic properties of a medium. This helps to
evaluate the decay rates and the scattering crossections of particles in
such a media. These results can be applied to check the abundances of light
elements in primordial nucleosynthesis, baryogenesis and leptogenesis. If
the background magnetic fields are also incorporated, then one can look for
applications to neutron stars, supernovae, red giants, and white dwarfs.

\bigskip

\bigskip \textbf{Figure caption}

Fig. 1\qquad A plot of wavefunction renormalization constant $Z_{2}^{-1}$ vs 
$T/m$ at two-loop level.

\textbf{References}

\begin{enumerate}
\item D. A. Kirzhnits, Field Theoretical Methods in Many Body Systems:
Permagon, Oxford, 1967; E. M.\ Lifshitz and L. P. Pitaevskii, Course on
Theoretical Physics- Physical Kinetics: Pergamon Press, New York.

\item T. Altherr and Kainulainen, Phys. Lett. B, 1991, 262: 79; T. Altherr,
Z. Phys. C, 1990, 47: 559 and Ann. Phys. (NY), 1991, 207: 374; E. Braaten,
Phys. Rev. Lett., 1991, 66: 2183; E. Braaten, Phys. Rev. Lett., 1991, 66:
1655; T. Altherr and U Kraemmer, Astroparticle Phys., 1992, 1: 133.

\item T. Matsubara, Prog. Theor. Phys., 1955, 14: 351.

\item J. Schwinger, J. Math. Phys., 1961, 2: 407.

\item R. Mills, Propagators for Many Particle Systems: Gordon and Breach,
New York, 1969.

\item L. V. Keldysh, Sov. Phys., 1964, 20: 1018.

\item C. Brouder, A. Frabetti hep-ph/0011161 and F. T. Brandt, Ashok Das, J.
Frenkel, Phys. Rev. D, 2002, 65: 085017.

\item See for example, H. Umezawa, H. Matsumoto and M. Tachiki, Thermo Field
Dynamics and Condensed States: North Holland, Amsterdam, 1982; A. Das,
Finite Temperature Field Theory: World Scientific, Singapore, 1997.

\item T. Kinoshita, J. Math. Phys., 1962, 3: 650.

\item J. F. Donoghue and B. R. Holstein, Phys. Rev. D28 (1983) 340; E.
Braaten and R. D. Pisarski, Nucl. Phys. B, 1990, 337: 567; R. Kobes, Phys.
Rev. D, 1990, 42: 562; L. Dolan and R. Jackiw, Phys. Rev. D, 1974, 9: 3320;
P. Landsman and Ch G. Weert, Phys. Rep., 1987, 145: 141.

\item K. Ahmed and Samina Saleem (Masood), Phys. Rev. D, 1987, 35: 1861;
ibid, 1987, 35: 4020; K. Ahmed and Samina S. Masood, Ann. Phys. (N.Y.),
1991, 207: 460; Samina Masood, Phys. Rev. D, 1991, 44: 3943.

\item Samina S. Masood, Phys. Rev. D, 1991, 44: 3943 ; Samina S. Masood,
Phys. Rev. D, 1993, 47: 648 ; ibid\ Phys. Rev. D, 1987, 36: 2602.

\item Mahnaz Qader (Haseeb), Samina S. Masood, and K. Ahmed, Phys. Rev. D,
1991, 44: 3322; ibid Phys. Rev. D, 1992, 46: 5633.

\item Samina S. Masood and Mahnaz Q. Haseeb, Int. J. of Mod. Phys. A, 2008,
23: 4709.

\item See for example: C. Itzykson and J. B. Zuber, Quantum Field Theory: \
McGraw- Hill Inc., 1990.
\end{enumerate}

\end{document}